\documentclass[12pt]{article}
\usepackage{epsfig}
\usepackage{cite}
\usepackage{bm}
\usepackage{amsmath}
\usepackage{graphics}

\usepackage[english]{babel}
\begin{document}


\vspace{1cm}
\title{\textbf{Transverse Momentum in Semi-inclusive \\ 
Deep Inelastic Scattering}}
\author{
\vspace{0.5cm}\\
\textbf{Federico Alberto Ceccopieri}
\vspace{0.3cm}\\
\textsl{Dipartimento di Fisica, Universit\'a di Parma,}\\
\textsl{Viale delle Scienze, Campus Sud, 43100 Parma, Italy}
\vspace{1cm}\\
\textbf{Luca Trentadue}
\vspace{0.3cm}\\
\textsl{Dipartimento di Fisica, Universit\'a di Parma,}\\
\textsl{INFN Gruppo Collegato di Parma,}\\
\textsl{Viale delle Scienze, Campus Sud, 43100 Parma, Italy}}
\date{}

\maketitle
\thispagestyle{empty}
\vspace{0.5cm}
\begin{center}
\large
Abstract
\vspace{0.5cm}\\
\end{center}
\normalsize
Within the framework of perturbative Quantum Chromodynamics
we derive transverse momentum dependent distributions describing   
both current and target fragmentation in semi-inclusive Deep Inelastic
Scattering. We present, to leading logarithmic accuracy,  the corresponding cross-sections
describing final state hadrons on the whole phase space.
Phenomenological implications and further developments are briefly discussed. 


\newpage

\begin{center}
\large{\bf{I. Introduction}}
\end{center}

The predictivity of perturbative Quantum Chromodynamics (pQCD) relies upon the factorization 
of hadronic cross-sections into perturbative process dependent coefficient functions, 
universal perturbative evolution equations and non-perturbative process 
independent densities\cite{fproof}. 
In presence of a hard scale, which justifies the perturbative approach due to asympotic freedom, 
the scale dependence of densities is predicted through 
renormalization group equations, \textsl{i.e.} Altarelli-Parisi (AP) evolution equations\cite{DGLAP}.
Inclusive deep inelastic lepton-hadron scattering (DIS) has been investigated since a long time
and in great detail.  
The evaluations of the splitting functions has been recently performed up to 
$\mathcal{O}(\alpha_s^3)$ in perturbation theory\cite{VVM}, further constraining 
non-perturbative dynamics. 
The evolution of initial state partons in terms  of longitudinal and transverse
momenta has been also considered within pQCD 
in Refs.\cite{kimber2,kimber,kwiechinski}. 
In semi-inclusive Deep Inelastic Scattering (SIDIS) processes, at variance with inclusive DIS, 
one hadron is detected in the final state, \textrm{$l+P\rightarrow l+h+X$}.
In this case, on the contrary, an equally 
accurate theoretical description has not yet been developed.
The additional hadronic degrees of freedom require a more detailed description of 
parton dynamics. 
Within the usual pQCD-improved parton model approach to SIDIS\cite{SIDIS_start,SIDIS_Cij},
one deals only with current fragmented hadrons.
However another distinct issue enters the perturbative description.
Both the struck parton and spectators do evolve according to the
hard scale governing the process\cite{Trentadue_Veneziano}. As a result also target
fragmentation has to be included to describe final state hadrons.
Evolution is predictable in terms of new functions dubbed \textsl{fracture functions},
whose factorization can been proven in Ref.\cite{Fact_M}.
An explicit evalution of the single-particle cross section at one loop  
has been given in Ref.\cite{Graudenz}.

In the pQCD-improved parton model detected hadrons are espected
to have a sizeable transverse momentum $\bm{P_{h \perp}}$, 
as a result of perturbative evolution in terms of hard partons emission. 
We will show that it is possible to reformulate evolution equations in order
to include transverse degrees of freedom since such a dependence is fully predictable within pQCD. 
 
The aim of this work is twofold. First we derive transverse momentum dependent (TMD)
evolution equations which enter SIDIS cross-sections in the current fragmentation region. 
We then extend this treatment to distributions in the target fragmentation region via fracture functions. 
As a result of these generalization, the combined  SIDIS cross section is presented 
descibing  hadron production on the whole phase space.


\vspace{0.5cm}
\begin{center}
\large{\bf{II. Transverse evolution equations and kinematics}}
\end{center}

Evolution equations to leading logarithmic accuracy (LLA) resum contributions due
to collinear partons emission. In the time-like case the evolution equation is:
\begin{equation}
\label{dglap}
Q^2 \frac{\partial \mathcal{D}_{i}^{h}(z_h,Q^2)}{\partial Q^2}=\frac{\alpha_s(Q^2)}{2\pi}\int_{z_h}^1 
\frac{du}{u} \,P_{ij}(u,\alpha_s(Q^2)) \,\mathcal{D}_{j}^{h}\Big(\frac{z_h}{u},Q^2\Big),
\end{equation}
where the fragmentation function $\mathcal{D}_{i}^{h}(z_h)$ represents the probability 
to find, at a scale $Q^2$, a given hadron $h$ with momentum fraction $z_h$ of its parent parton $i$.
$P(u)$ are the time-like splitting functions which give the probabilities
to find a parton of type $j$ inside a parton of type $i$ and can be 
expressed as a power series in the  strong running coupling $\alpha_s(Q^2)$ .
Ordinary evolution equations, eq.(\ref{dglap}), contain only longitudinal 
degrees of freedom of partons inside hadrons although,  
at each branching, the emitting parton acquires a transverse momentum 
relative to its initial direction. 
Transverse momentum dependent (TMD) evolution equations were 
first derived in Ref.\cite{BCM} for fragmentation functions in the time-like region:
\begin{eqnarray}
\label{dglap_TMD_time}
Q^2 \frac{\partial \mathcal{D}_{i}^{h}(z_h,Q^2,\boldsymbol{p_{\perp}})}{\partial Q^2}&=&
\frac{\alpha_s(Q^2)}{2\pi}\int_{z_h}^1 \frac{du}{u} 
P_{ij}(u,\alpha_s(Q^2))\cdot\nonumber\\
&&\cdot \frac{d^2 \boldsymbol{q_{\perp}}}{\pi}\,\delta(\,u(1-u)Q^2-q^2_{\perp})\,
\mathcal{D}_{j}^{h}\Big(\frac{z_h}{u},Q^2,\boldsymbol{p_{\perp}}-\frac{z_h}{u} \bm{q_{\perp}} \Big).
\end{eqnarray}
Single particle distributions $\mathcal{D}_{i}^{h}(z_h)$ in eq.(\ref{dglap_TMD_time}) give
the probability to find, at a given scale $Q^2$, the hadron $h$ in the parent parton $i$
with longitudinal momentum fraction $z_h$ and transverse momentum $\bm{p}_{\perp}$ with respect to it. 
The $P(u)$ splitting functions are the ordinary splitting functions
as in eq.(\ref{dglap}) and flavour indices are understood as in the inclusive case. 
\begin{figure}[h]
\begin{center}
\epsfig{file=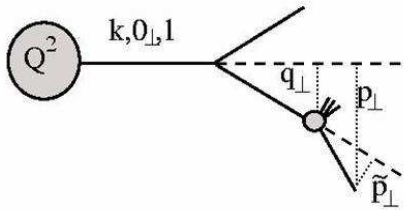,width=6cm,height=3.5cm,angle=0}
\caption{\small{a parton with momentum $k$ emerges from a hard process with virtual mass $k^2=Q^2>0$ 
and then evolves into a quasi-real parton with transverse momentum $\bm{p}_{\perp}$. 
The vertex is associated with the time-like splitting functions $P(u)$.
The small blob symbolizes resummation of ladder diagrams in LLA. }}
\end{center}
\label{fig:tmdtime}
\end{figure}
\normalsize
Let us discuss the kinematics structure of eq.(\ref{dglap_TMD_time})\,.
Consider an outgoing parton $k$ emerging from an hard collision with 
virtuality  $k^2=Q^2>0$, assigned to have zero transverse momentum ($\bm{k}_{\perp}=\bm{0}_{\perp}$) and 
unitary longitudinal momentum fraction, as displayed in Fig.(1)\,. 
It subsequently branches, with probability $P(u)$, into a couple of partons,
$q$ and $q'$, with a fractional momentum $u$ and $1-u$ of $k$ respectively and 
transverse momentum $\bm{q}_{\perp}=-\bm{q'}_{\perp}$ relative to $\bm{k}$. 
The parton $q$ then non-perturbatively hadronizes generating the final hadron $h$ 
with a fractional momentum $z_h$ and transverse momentum $\bm{p}_{\perp}$ and $\bm{\widetilde{p}}_{\perp}$
relative to $\bm{k}$ and $\bm{q}$ respectively.  
We thus derive the following constraints:
\begin{eqnarray}
\label{boost+flow}
\tilde{\bm{p}}_{\perp}&=&\bm{p}_{\perp}-\frac{z_h}{u}\,\bm{q}_{\perp} \, , \\
q^2_{\perp}&=&u\,(1-u)\,Q^2 \,. 
\end{eqnarray}
Eq.(\ref{boost+flow}) takes into account the Lorentz boost of transverse momenta $\widetilde{\bm{p}}_{\perp}$
from the $q$-frame to the $p$-frame. The second equation follows by 
imposing the virtuality flow of time-like branching. 
These relations directly enter eq.(\ref{dglap_TMD_time}), respectively as argument of the 
distribution $\mathcal{D}_{j}^{h}$ and of the invariant-mass conserving $\delta$-function.  
The unintegrated densities fulfil the normalization:
\begin{equation}
\label{timelike_norm}
\int d^2 \bm{p}_{\perp} \mathcal{D}_{i}^{h}(z_h,Q^2,\bm{p}_{\perp})=\mathcal{D}_{i}^{h}(z_h,Q^2)\,,
\end{equation}
since the boost in eq.(\ref{boost+flow}) is linear in the transverse variables, \textsl{i.e.}
the Jacobian is:
\begin{equation}
\label{jacobian}
\det \Big( \frac{d^2 \bm{p}_{\perp}}{d^2 \widetilde{\bm{p}}_{\perp}} \Big)=1\,.
\end{equation} 
This property garantees that we can recover inclusive distributions, eq.(\ref{dglap}), starting 
from less inclusive ones. The opposite statement is not valid since eq.(\ref{dglap_TMD_time}) 
contains new physical information. 

In order to obtain a complete description of semi-inclusive cross-sections 
we need the space-like version of eq.(\ref{dglap_TMD_time}). On a general ground
we may expect that the infrared structure of space-like evolution equations 
is the same as the time-like one since it depends only upon the dynamics of the underlying gauge
theory, the only changes being in the kinematics.
In analogy to the time-like case we consider now a initial state parton $p$
in a incoming proton $P$ which undergoes a hard collision, 
the reference frame being aligned along the incoming proton axis, as in Fig.(2).
\begin{figure}
\begin{center}
\label{fig_tmd_space}
\epsfig{file=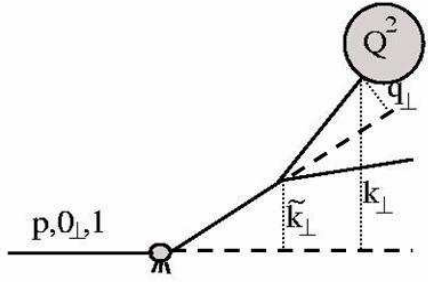,width=6cm,height=4cm,angle=0}
\caption{\small{a quasi-real parton with momentum $p$ perturbatively evolves towards the hard vertex
acquiring a virtual mass $p^2=Q^2<0$.  
The vertex is associated with the space-like splitting functions $P(u)$.
The small blob symbolizes resummation of ladder diagrams in LLA. }}
\end{center}
\end{figure}
\normalsize
The boost of transverse momentum and the invariant mass-conserving constraint are in this case:
\begin{eqnarray}
\label{boost+flow2}
\tilde{\bm{k}}_{\perp}&=&\frac{\bm{k}_{\perp}-\bm{q}_{\perp}}{u}, \\
q^2_{\perp}&=&(1-u)\,Q^2 \,. 
\end{eqnarray}
We thus generalize eq.(\ref{dglap_TMD_time}) to the space-like case:
\begin{eqnarray}
\label{dglap_TMD_space}
Q^2 \frac{\partial \mathcal{F}_{P}^{i}(x_B,Q^2,\boldsymbol{k_{\perp}})}{\partial Q^2}
&=&\frac{\alpha_s(Q^2)}{2\pi}\int_{x_B}^1 \frac{du}{u^3} 
P^{i}_{j}(u,\alpha_s(Q^2))\cdot\nonumber\\
&&\cdot \frac{d^2 \boldsymbol{q_{\perp}}}{\pi}\,\delta(\,(1-u)Q^2-q^2_{\perp})
\,\mathcal{F}_{P}^{j}\Big(\frac{x_B}{u},Q^2, \frac{\bm{k}_{\perp}-\bm{q}_{\perp}}{u} \Big)\,.
\end{eqnarray}
The \textsl{constructive} proof of eq.(\ref{dglap_TMD_space}) will be given elsewhere\cite{newpaper}.
The unintegrated distributions fulfil a condition
analogous to the one in eq.(\ref{timelike_norm}), \textsl{i.e.} :
\begin{equation}
\label{spacelike_norm}
\int d^2 \bm{k}_{\perp} \mathcal{F}_{P}^{i}(x_B,Q^2,\bm{k}_{\perp})=\mathcal{F}_{P}^{i}(x_B,Q^2)\,.
\end{equation}
We note that the inclusion of transverse momenta do not affect longitudinal degrees of freedom since partons 
always degrade their fractional momenta in the perturbative branching processes.
The different Lorentz structure in the transverse arguments of the
parton distribution functions $\mathcal{F}_ {P}^{i}$ arises from the different structure of the 
Bethe-Salpeter ladder used to derive the evolution equations\cite{newpaper}.

\vspace{0.5cm}
\begin{center}
\large{\bf{III. Transverse momenta in target fragmentation region}}
\end{center}

Current and target hadron production mechanisms cannot be separated since 
hadrons produced by current fragmentation may go in to the target remnant direction and \textsl{vice versa}.
In these configurations new infrared singularities appear 
which cannot be reabsorbed through the standard renormalization procedure into parton distribution functions and 
fragmentation functions. It has been shown\cite{Trentadue_Veneziano,Fact_M,Graudenz}
that the cross-section can be renormalized by introducing new non-perturbative 
objects, fracture functions indicated by $\mathcal{M}^{i}_{P,h}(x,z,Q^2)$. 
These functions express the conditional probability 
of finding, at a scale $Q^2$, a parton $i$ with momentum fraction $x$ in a proton $P$
while a hadron $h$ with momentum fraction $z$ is detected. 
Fracture functions obey non-homogeneous evolution equations\cite{Trentadue_Veneziano}: 
\begin{eqnarray}
\label{M-evo_long}
Q^2 \frac{\partial \mathcal{M}^{i}_{P,h}(x,z,Q^2)}{\partial Q^2}=
\frac{\alpha_s(Q^2)}{2\pi}\int_{\frac{x}{1-z}}^{1}
\frac{du}{u}\,P^{i}_{j}(u)\mathcal{M}^{j}_{P,h}\Big(\frac{x}{u},z,Q^2\Big)+&&\nonumber\\
+ \frac{\alpha_s(Q^2)}{2\pi}
\int_{x}^{\frac{x}{x+z}}\frac{du}{x(1-u)} \hat{P}^{i,l}_{j}(u)
\mathcal{F}_{P}^{j}\Big( \frac{x}{u},Q^2\Big)
\mathcal{D}_{l}^{h} \Big( \frac{zu}{x(1-u)} ,Q^2 \Big).&&
\end{eqnarray}
The first term in the above equation (see Fig.(3a)) describes the evolution of the active parton $j$ while 
the hadron $h$ is detected. The second term (see Fig.(3b)) takes into account the
production of a hadron $h$ by a time-like cascade initiated by the active parton $j$.
The $\hat{P}^{i,l}_{j}(u)$ represent the unsubtracted Altarelli-Parisi splitting functions\cite{unregAP}.
Since perturbative evolution is at work even in target fragmentation region, we 
expect that a non negligible amount of transverse momentum is produced there. 
We thus generalize these distributions to contain also transverse degrees of freedom. The fracture functions 
$\mathcal{M}^{i}_{P,h} (x,\bm{k}_{\perp},z,\bm{p}_{\perp},Q^2)$ give the conditional probability 
to find in a proton $P$, at a scale $Q^2$, a parton with momentum fraction $x$ and transverse momentum 
$\bm{k}_{\perp}$ while a hadron $h$, with momentum fraction $z$ 
and transverse momentum $\bm{p}_{\perp}$, is detected. 
Under these assumptions the following evolution equations can thus be derived\cite{newpaper}:
\begin{eqnarray}
\label{M-evo_long+tra}
&&Q^2\frac{\partial \mathcal{M}^{i}_{P,h}
(x,\bm{k}_{\perp},z,\bm{p}_{\perp},Q^2)}{\partial Q^2}=
\frac{\alpha_s(Q^2)}{2\pi} \Bigg\{ \int_{\frac{x}{1-z}}^{1} \frac{du}{u^3} \,P^{i}_{j}(u)\, 
\int\frac{d^2 \bm{q}_{\perp}}{\pi}\, \delta(\,(1-u)Q^2-q_{\perp}^2)\cdot\nonumber\\
&&\cdot\mathcal{M}^{j}_{P,h}\Big(Q^2,\frac{x}{u},\frac{\bm{k}_{\perp}-\bm{q}_{\perp}}{u},
z,\bm{p}_{\perp} \Big)+ \int_{x}^{\frac{x}{x+z}} \frac{du}{x(1-u)u^2} \hat{P}^{i,l}_{j}(u)
\frac{d^2 \bm{q}_{\perp}}{\pi}\,\delta(\,(1-u)Q^2-q_{\perp}^2) \cdot \nonumber\\
&& \cdot \mathcal{F}_{P}^{j} 
\Big(\frac{x}{u},\frac{\bm{k}_{\perp}-\bm{q}_{\perp}}{u},Q^2 \Big)\,
\mathcal{D}_{l}^{h}\Big(\frac{zu}{x(1-u)},\bm{p}_{\perp}-\frac{zu}{x(1-u)}\,
\bm{q}_{\perp},Q^2 \Big)\Bigg\} \,. 
\end{eqnarray}  
As in the longitudinal case two terms contribute to the evolution of
TMD fracture functions as displayed in Fig.(3). The homogeneous one has a pure non-perturbative 
nature since involves the fragmentation of the proton remnants into the hadron $h$. 
The inhomogeneous one takes into account the production of the hadron $h$ from a time-like cascade of 
parton $j$ and thus is dubbed \textsl{perturbative}.
\begin{figure}[h]
\begin{center}
\label{TMD_fracture_evo}
\epsfig{file=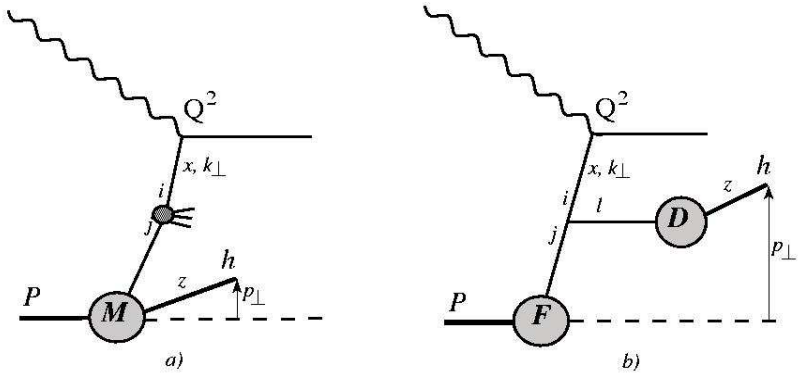,width=11cm,height=5cm,angle=0}
\caption{\small{ Evolution of fracture functions $M$:   
a) homogeneous term; b) inhomogeneous one.}}
\end{center}
\end{figure}
\normalsize
The TMD fracture functions fulfil the normalization condition:
\begin{equation}
\label{M_norm}
\int d^2 \bm{k}_{\perp}\int d^2 \bm{p}_{\perp} \mathcal{M}^{i}_{P,h}
(x,\bm{k}_{\perp},z,\bm{p}_{\perp},Q^2) =  \mathcal{M}^{i}_{P,h}
(x,z,Q^2) \,,
\end{equation}
as direct consequence of the kinematics of both terms in the evolution equations, eq.(\ref{M-evo_long+tra}).
The proof of factorization, \textsl{i.e.} that all singularities 
occuring in the target remnant direction can be properly renormalized by 
the less inclusive TMD quantity  $\mathcal{M}^{i}_{P,h}
(x,\bm{k}_{\perp},z,\bm{p}_{\perp},Q^2)$, is still lacking at present.
In the following we assume such a factorization to hold. 

\vspace{0.5cm}
\begin{center}
\large{\bf{IV. Solutions}}
\end{center}

TMD evolution equations (\ref{dglap_TMD_time}) and (\ref{dglap_TMD_space})  
can only be approximately diagonalized by the joint Fourier-Mellin transform
\begin{equation}
\label{Fourier-Mellin}
\mathcal{D}_n(\bm{b},Q^2)=\int d^2 \bm{p}_{\perp}\int_0^1 \makebox{e}^{\frac{-i\,\bm{b}\cdot\,\bm{p}_{\perp}}{z}}
dz \; z^n \,\mathcal{D}(z,Q^2,\bm{p}_{\perp}) \, , 
\end{equation}
where $\bm{b}$ is the transverse momentum Fourier-coniugated variable\cite{PP}. 
An exact diagonalization is prevented by the kinematics structure 
of the distribution under integral in eq.(\ref{dglap_TMD_time})
since it combines longitudinal momentum fractions with transverse momenta. 
Such an exact diagonalization can be however obtained in the soft limit,
\textsl{i.e.} when the variables $x$ or $z$ approach the edge of phase space. 
In this case for the non-singlet time-like unintegrated fragmentation functions
the solution reads\cite{BCM} :
\begin{equation}
\label{soft_solution}
\mathcal{D}(z,Q^2,\bm{p}_{\perp})=\mathcal{D}(z,Q^2)\;\mathcal{G}(Q_0^2,Q^2,z,\bm{p}_{\perp})\,,
\end{equation}
where  the scale $Q_0^2$ sets the upper limit of the non-perturbative regime.
We recall that in this limit the convergence of the perturbative 
series can be further improved by taking into account soft gluon radiation enhancements. 
The form factor $\mathcal{G}$ can therefore be computed to leading logarithmic accuracy\cite{CG,ABCMV,PP}
by simply demanding that $\alpha_s(Q^2)\rightarrow\alpha_s(\bm{p}_{\perp}^2)$. 
The expression to next-to-leading logarithmic accuracy has been given in Ref.\cite{kodaira}. 
These issues have been recently specialized to the case of SIDIS
processes in the current fragmentation region in Ref.\cite{resummation_low_pt}. 
Away from the soft limit, the factorized structure of the solution, eq.(\ref{soft_solution}), 
is not automatically preserved. In this case numerical methods have shown to be 
useful in order to solve the equations in the $x,z \in \mathcal{O}(1)$ range\cite{Jones,newpaper}
and to extend the solutions to be valid in the flavour mixing sector.
As in the longitudinal case, distributions at a
scale $Q^2>Q_0^2$ are known if we provide a non-perturbative input density at some arbitrary scale $Q_0^2$.
We assume as initial condition the usual longitudinal fragmentation
distribution times a $z$-independent, flavour independent factor:
\begin{equation}
\label{ansatz}
\mathcal{D}^{h}_{i}(z,Q_0^2,\bm{p}_{\perp})= \mathcal{D}^{h}_{i}(z,Q_0^2)
\;\frac{\makebox{e}^{-\bm{p}_{\perp}^2/<\bm{p}_{\perp}^2>}}{\pi<\bm{p}_{\perp}^2>} \,.
\end{equation}
The gaussian $\bm{p}_{\perp}$-distribution
is used to model partons intrinsic momenta inside hadrons\cite{Cahn}.
This issue has also been considered in Refs.\cite{Chay,anselmino}.  
\begin{figure}[h]
\begin{center}
\label{H2def}
\epsfig{file=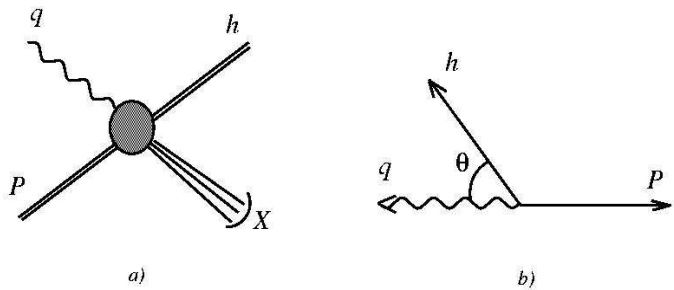,width=9cm,height=4cm,angle=0}
\caption{\small{a) semi-inclusive DIS \textrm{$l+P\rightarrow l+h+X$} process; b) 
kinematics of the reaction in the photon-proton center of mass frame.}}
\end{center}
\end{figure}
\normalsize
We now discuss  the phenomenological implications of these results. 
According to Ref.\cite{h2_mulders}, 
single hadron cross-sections are usually 
parametrized in terms of four independent structure functions 
$H_{i=1,..,4}(x_B,z_h,Q^2,\bm{P}_{h\perp})$.
The hadronic invariants of SIDIS processes are constructed by using 
external particles four-momenta, as displayed in Fig.(4a), and therefore 
the structure functions $H_{i}$ are supposed to 
describe both current and target hadron fragmentation mechanisms. 

Once TMD evolution equations are solved,
these predictions can be compared with semi-inclusive DIS data, as for the longitudinal case, provided that 
a factorization theorem holds even for TMD distributions. 
Such a theorem has been shown to hold in the current fragmentation
region for the structure function $H_2$ in Ref.\cite{Ji} : 
\begin{eqnarray} 
\label{kt_sidis_fact}
&&H_2(x_B,z_h,\bm{P}_{h\perp},Q^2)=\sum_{i=q,\,\bar{q}}
e_q^2\int d^2\bm{k}_{\perp} d^2\bm{p}_{\perp} \delta^{(2)} 
(z_h \bm{k}_{\perp}+\bm{p}_{\perp} - \bm{P}_{h\perp})\cdot\nonumber\\
&&\quad\quad\quad\quad\quad\quad\quad\quad\quad
\cdot\;\mathcal{F}_{P}^{i}(x_B,\mu^2_F,\bm{k}_{\perp},) \; \mathcal{D}^{h}_{i}(z_h,\mu^2_D,\bm{p}_{\perp}) 
\; C(Q^2,\mu^2_F,\mu^2_D) \,,
\end{eqnarray}
where the standard SIDIS variables are defined as:
\begin{equation}
z_h=\frac{P\cdot P_h}{P\cdot q}, \; \; \; x_B=\frac{Q^2}{2 P \cdot q} \,,
\end{equation}
and $\mu^2_{F}$ and  $\mu^2_{D}$ are the factorization scales.  
The above results are accurate up to powers in $(P_{h\perp}^2/Q^2)^n$ for soft 
transverse momenta $P_{h\perp}\simeq\Lambda_{QCD}$.
Evolution equations for $\mathcal{F}$ and $\mathcal{D}$ are given in 
eqs.(\ref{dglap_TMD_time}) and (\ref{dglap_TMD_space}).
The factor $C$ is the process-dependent hard
coefficient function computable in perturbative QCD and to LLA we can set $\makebox{C=1}$.
Provided that factorization holds for the TMD fracture functions,  
we may add, according to eq.(\ref{M-evo_long+tra}), their contributions to $H_2$:
\begin{eqnarray}
\label{SIDIS_cross_section}
&&H_2(x_B,z_h,\bm{P}_{h\perp},Q^2)=\sum_{i=q,\bar{q}}
e_q^2\int d^2\bm{k}_{\perp} d^2\bm{p}_{\perp} \Big\{\delta^2 
(z_h \bm{k}_{\perp}+\bm{p}_{\perp} - \bm{P}_{h\perp})\cdot\\
&&\cdot\mathcal{F}_{P}^{i}(x_B,Q^2,\bm{k}_{\perp}) \, 
\mathcal{D}_{i}^{h}(z_h,Q^2,\bm{p}_{\perp})A(0)
+(1-x_B) \mathcal{M}^{i}_{P,h} (x_B,\bm{k}_{\perp},z,\bm{p}_{\perp},Q^2)
\,\delta^2 (\bm{p}_{\perp} - \bm{P}_{h\perp})A(1)\Big\}\nonumber
\end{eqnarray} 
where we have identified all the three factorization scales with the hard scale,  
$Q^2=\mu^2_F=\mu^2_D=\mu^2_M$.
Although formally the two contributions are simply added in eq.(\ref{SIDIS_cross_section}), 
at LLA and in photon-proton center of mass frame, 
the produced hadrons are mainly distributed in two opposite emispheres. 
Target fragmentated hadrons are produced mainly in the $\theta=\pi$ direction 
while current fragmented hadrons mainly along $\theta=0$ direction. 
Here $\theta$ is the angle of the produced hadron $h$ with respect to the photon direction,
as shown in Fig.(4b).
In order to keep track of the emission angle of the detected hadron $h$, we supplement 
current and target framentation terms in
eq.(\ref{SIDIS_cross_section}) with an angular distribution $A(v)$\cite{Graudenz}.
The angular and energy variables $v$ and $z$ are defined as: 
\begin{equation}
\label{z_def_cms}
z=\frac{E_h}{E_p(1-x_B)}, \quad \quad v=\frac{1-\cos\theta}{2},  \quad
\quad z_h= z\,v\,. 
\end{equation}
In eq.(\ref{z_def_cms}), $E_h$ and $E_p$ denote respectively 
the energies of the detected hadron and of the incoming proton 
in the photon-proton center of mass frame. The variables $z$ and $v$ are a useful frame-dependent 
representation for the hadronic invariant $z_h$ in two respects: $z$
reduces to $z_h$ in the current fragmentation region so that we recover the standard definitions, 
while for low $z_h$-values we can distinguish soft hadrons ($z\rightarrow 0$) from the ones
produced in the target remnant direction ($\theta\rightarrow \pi$).
Since to LLA all sources of transverse momenta contributing to
$\bm{P}_{h\perp}$ have been taken into account we may 
pictorially represent eq.(\ref{SIDIS_cross_section}) as in Fig.(5).             
\begin{figure}[ht]
\begin{center}
\label{kt_sources_target}
\epsfig{file=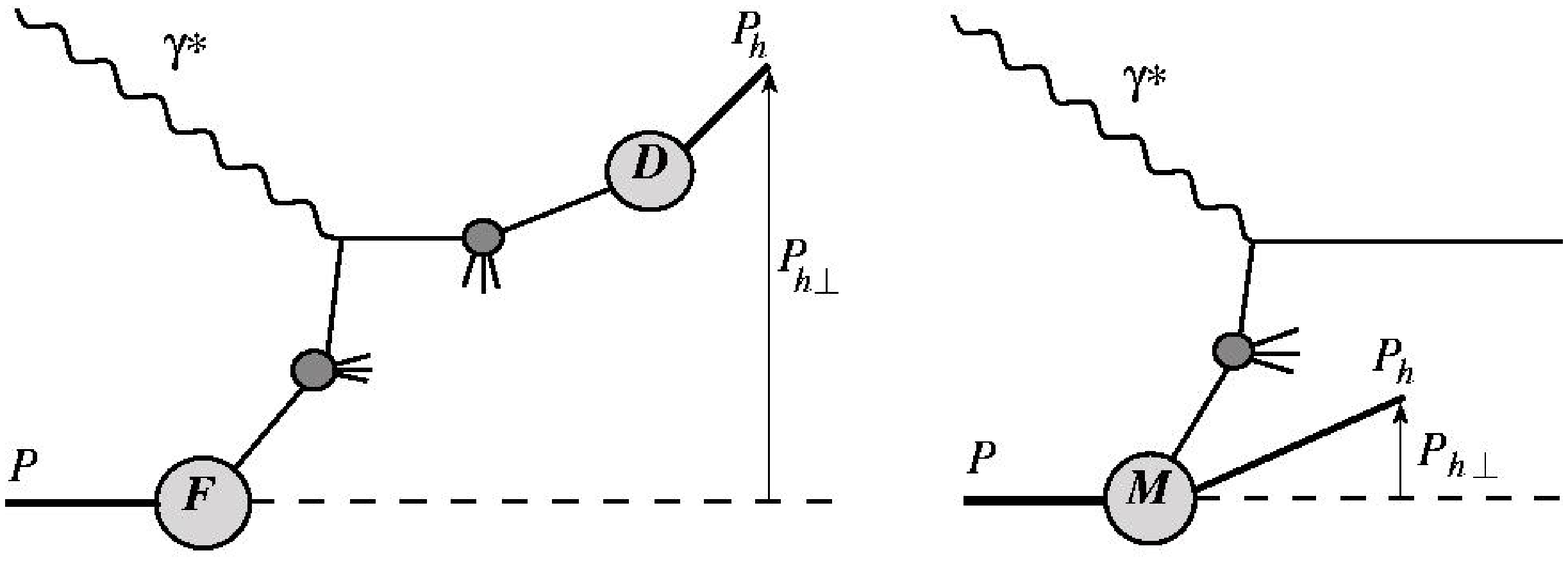,width=10cm,height=4cm,angle=0}
\caption{\small{ Sources of transverse momentum 
in the current (left) and in the target (right) fragmentation region
in SIDIS. Dark blobs symbolize hard partons emission.
Transverse momentum $\bm{P}_{h\perp}$ of the detected hadron $h$ is
also indicated.
$F$, $D$ and $M$ represent parton distribution, fragmentation and fracture functions respectively.}}
\end{center}
\end{figure}
\normalsize

\vspace{0.5cm}
\begin{center}
\large{\bf{V. Conclusions and perspectives}}
\end{center}
In this work we have extended ordinary distributions to
include transverse degrees of freedom both in the current and in the target 
fragmentation region of semi-inclusive DIS. 
As long as a factorization theorem holds for transverse  momentum dependent fracture functions,
the semi-inclusive cross-sections are thus predictable on the whole phase space of the detected hadrons. 
Although this extension may have its own theoretical relevance, on the 
phenomenological side it also improves our knowledge of both
the perturbative and non-perturbative dynamics of partons.
Evolution equations (\ref{dglap_TMD_time}), (\ref{dglap_TMD_space}) and (\ref{M-evo_long+tra}) 
can be straightforwardly extended also to the polarized case. In the target 
fragmentation region the evolution equations for longitudinal polarized fracture
functions have been derived in Ref.\cite{sassot_polarized}.
We stress that the accuracy of
eqs.(\ref{dglap_TMD_time}), (\ref{dglap_TMD_space}) and (\ref{M-evo_long+tra}) 
is set by the accuracy of the Altarelli-Parisi splitting functions. Splitting functions for partons 
and fragmentation functions are well known.
They have been recently calculated in the target region at two loop level in
Ref.\cite{sassot_2loop}. Therefore the evolution equations in the current and in the
target fragmentation region can be set as to allow the analysis of the semi-inclusive cross sections
to the same level of accuracy. Detailed derivation of the results presented in this work and of the phenomenological 
implications are postponed to a forthcoming paper\cite{newpaper}.

\vspace{0.5cm}
\begin{center}
\large{\bf{VI. Acknowledgments}}
\end{center}

We would like to thank Mikhail Osipenko for fruitful discussions on
the role of transverse momentum distributions in the data analysis of SIDIS processes.
One of us, F.A.C, thanks the Dipartimento di Fisica dell' Universit\'a di Parma for the ospitality 
given to him.  

\newpage
\begin{center}

\end{center}
\end{document}